%% file: samplepaper.tex
\documentclass[runningheads]{llncs}
\usepackage{url}
\usepackage{graphicx}
\usepackage{listings}
\usepackage{setspace}
\usepackage{pgfplots}
\pgfplotsset{compat=1.18}
\usepackage[T1]{fontenc}
\usepackage{xcolor}
\definecolor{arm-orange}{RGB}{255,107,0}
\definecolor{arm-light-blue}{RGB}{0,193,222}
\definecolor{arm-light-gray}{RGB}{229,236,235}
\definecolor{arm-yellow}{RGB}{255,199,0}
\definecolor{arm-blue}{RGB}{0,145,189}
\definecolor{arm-dark-gray}{RGB}{125,134,140}
\definecolor{arm-green}{RGB}{149,214,0}
\definecolor{arm-dark-blue}{RGB}{0,43,73}
\definecolor{arm-black}{RGB}{51,62,72}
\definecolor{code-comment-green}{rgb}{0,0.6,0}
\definecolor{code-string}{rgb}{0.58,0,0.82}
\usepackage{tikz} 
\usetikzlibrary{calc,fit}
\usepackage{appendix}
\usepackage[inline]{enumitem}
\usepackage{booktabs}
\usepackage{multirow}
\usepackage[backend=biber,style=numeric]{biblatex}
\begin{document}
\title{Verifying components of Arm\textsuperscript{\textregistered} Confidential Computing Architecture with ESBMC}
\titlerunning{Verifying Arm\textsuperscript{\textregistered} RMM with ESBMC}
\author{Tong Wu\inst{1}\orcidID{0000-0002-0986-4150} \and
Shale Xiong\inst{2}\orcidID{0000-0001-9312-195X} \and
Edoardo Manino\inst{1}\orcidID{0000-0001-9312-195X} \and
Gareth Stockwell\inst{2}\orcidID{0009-0004-1773-2846} \and
Lucas C. Cordeiro\inst{1}\orcidID{0000-0002-6235-4272}}
\authorrunning{T. Wu et al.}
\institute{The University of Manchester, Manchester, UK \\
\email{tong.wu-11@postgrad.manchester.ac.uk}\\
\email{\{edoardo.manino,lucas.cordeiro\}@manchester.ac.uk}\and
Arm\textsuperscript{\textregistered}\\
\email{\{shale.xiong,gareth.stockwell\}@arm.com}}
\maketitle              % typeset the header of the contribution
\begin{abstract}
\emph{Realm Management Monitor} (RMM) is an essential firmware component within the recent Arm \emph{Confidential Computing Architecture} (Arm CCA). Previous work applies formal techniques to verify the speciﬁcation and prototype reference implementation of RMM. However, relying solely on a single verification tool may lead to the oversight of certain bugs or vulnerabilities. This paper discusses the application of ESBMC, a state-of-the-art Satisfiability Modulo Theories (SMT)-based software model checker to further enhance RRM verification.
We demonstrate ESBMC's ability to precisely parse the source code and identify specification failures within a reasonable time frame. Moreover, we propose potential improvements for ESBMC to enhance its efficiency for industry engineers. This work contributes to exploring the capabilities of formal verification techniques in real-world scenarios and suggests avenues for further improvements to better meet industrial verification needs.

\end{abstract}
%
%
%--------------------------------------
\section{Introduction}
%--------------------------------------

The rise of Confidential Computing is driven by the need to secure computations in Cloud Computing, where sensitive information is delegated to a third party, posing potential security risks~\cite{SubashiniK11}. For example, hundreds of millions of Facebook user records were exposed on Amazon cloud server~\cite{silverstein2019hundreds}.

Confidential Computing technologies, including Arm \emph{Confidential Computing Architecture} (Arm CCA)~\cite{armcca}, introduce protected execution environments, for example \emph{Realms} in Arm CCA, to ensure confidentiality and integrity \textcolor{black}{of sensitive information.} In particular, Arm CCA allows multiple Realms to execute in parallel; to manage all instances, Arm CCA then introduces a new privileged firmware component called \emph{Realm Management Monitor} (RMM). RMM also provides a necessary interface to the outside, non-secure world. Arm is committed to specifying the behavior of RMM~\cite{arm2022realm} and providing an implementation in C, and the code is already open-sourced to the community~\cite {tfrmm}.

RMM is a critical component, and any end-user must trust it. Arm provides a specification~\cite{arm2022realm} for all the interfaces via pairs of pre/post conditions. To verify the correctness of the RMM implementation concerning the RMM speciﬁcation, formal methods such as interactive theorem provers and symbolic model checking can be applied to automate the verification workflow~\cite{varmcca}. Arm engineers automatically generate a verification harness from a machine-readable specification of the RMM by evaluating the pre- and post-conditions to constrain the inputs and outputs. This harness can be consumed by \textcolor{black}{a software verification tool such as} \emph{C Bounded Model Checker} (CBMC) \cite{amazon2022cbmcviewer,kroening2014cbmc}. Arm researchers and engineers are gradually deploying these auto-generated verification harnesses and CBMC in the CI/CD system. \textcolor{black}{Several} violations in the implementations detected by CBMC \textcolor{black}{have been} confirmed and fixed \textcolor{black}{by Arm engineers, thus} demonstrating the \textcolor{black}{value of software model checking} techniques~\cite{fox2023verification} \textcolor{black}{in strengthening safety-critical systems}.

Given the fact that there exist some violated properties in the \textcolor{black}{current} draft implementation \textcolor{black}{of the RMM}~\cite{fox2023verification}, we are keen to explore the following questions:
\begin{itemize}
\item[{$\bullet$}] Is the existing verification \textcolor{black}{enough to secure RMM?}
\item[{$\bullet$}] If not, \textcolor{black}{can} other state-of-the-art techniques \textcolor{black}{find additional violations?}
\end{itemize}

To explore these questions, we further apply ESBMC, an efficient model checker based on SMT theories that can automatically detect or prove the absence of runtime errors in software written in C/C++, Kotlin, Python, and Solidity~\cite{GadelhaMMC0N18, MenezesAFLMSSBGTKC24}. According to the recent Competitions on Software Verification (SV-COMP)~\cite{beyer2023svcomp, Beyer24}, ESBMC consistently outperforms CBMC in proving more safety properties while producing fewer incorrect results. However, there is little evidence of whether this superior performance could uncover more vulnerabilities in real-world low-level software systems. Through our exploration of ESBMC on some RMM verification cases, we achieved the following contribution:
\begin{itemize}
    \item [{$\bullet$}]We reproduced the same failures reported by CBMC, which were confirmed by Arm engineers previously.
    \item [{$\bullet$}]We identified inconsistent results from CBMC and reported them to the developers.
    \item [{$\bullet$}]\textcolor{black}{We found 23 new violations in the RMM code that only ESBMC can detect.}
    \item [{$\bullet$}]\textcolor{black}{We showed that the verification performance could be significantly improved by efficiently configuring the bounds for each loop in the program.} 
    \item [{$\bullet$}]We highlighted the challenge of checking multiple properties and contributed to its development within ESBMC. 
\end{itemize} 

%---------------------------------------------
\section{Background}
\subsection{Bounded Model Checking (BMC)}
%---------------------------------------------

SAT/SMT-based BMC is a formal verification technique to falsify or prove the correctness of finite-state systems~\cite{ClarkeBRZ01}. This method explores a system's state space up to a specified depth or bound $k$, searching for potential errors or violations within that limited scope. Then the program is encoded into an SAT/SMT formula, and it is satisfiable \textit{iff} there exists a counterexample that violates the property~\cite{biere2009bounded}. 

As a BMC tool, ESBMC implements various algorithms for bounded and incremental verification~\cite{AlhawiRGCF21,CordeiroFM12,GadelhaIC17}: \begin{enumerate*}[label=(\arabic*)] \item incremental BMC, which unrolls the loops and checks the property incrementally; \item \textit{k}-induction~\cite{gadelha2019esbmc} proof-rule algorithm, which aims to prove the safety based on induction; and \item default BMC approach that bounds the program up to a given depth and call the SMT solver once to solve the entire SMT formula. 
\end{enumerate*}

Similar to CBMC, ESBMC takes an RMM test case as input and checks the safety of the properties in the given program. In particular, ESBMC can check for user-defined assertions, pointer safety, buffer overflow, data races, and so on through configurations. The ESBMC documentation about the supported safety/security properties, underlying verification algorithms, and integrated SMT solvers is available online.\footnote{\url{https://ssvlab.github.io/esbmc/documentation.html}} 

%----------------------------------------------------
\subsection{Realm Management Monitor}
%----------------------------------------------------

\begin{figure}[t]
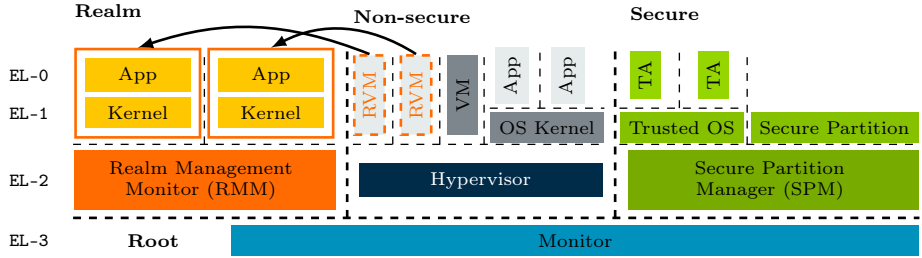

  \centering
  \include{cca-diagram.tex}
  \caption{An architecture of an Arm CCA~\cite{fox2023verification}. The
physical memory space is separated into four worlds:
Non-secure, Secure, Root and the Realm.} %\textcolor{red}{LC: can you briefly describe this figure?}}
  \label{fig:cca}
\end{figure}

Fig.~\ref{fig:cca} illustrates the architecture of Arm CCA; more can be learned from~\cite{armcca}. Executions Arm's \emph{Processing Element} (PE) is associated with levels EL0 to EL3, where user-space executes at EL0 and most privileged low-level firmware, or monitor, executes at EL3. The physical memory space also separates into four parts, or four worlds, namely, \emph{Non-secure} World, \emph{Secure} World, \emph{Root} World, and the newly-introduced \emph{Realm}. In the \emph{Realm} World, an end-user application, referred to as a realm, is executed at the level of EL0 and EL1. To administer Realms, Arm CCA introduces a new privileged firmware component called the \emph{Realm Management Monitor} (RMM), executing at EL2 in the Realm world, which acts as a separation kernel isolating Realms from each other. Also, RMM provides interfaces to the Non-secure World for managing and scheduling realms indirectly.

\begin{figure}[t]
  \centering
  \includegraphics[width=1.0\textwidth]{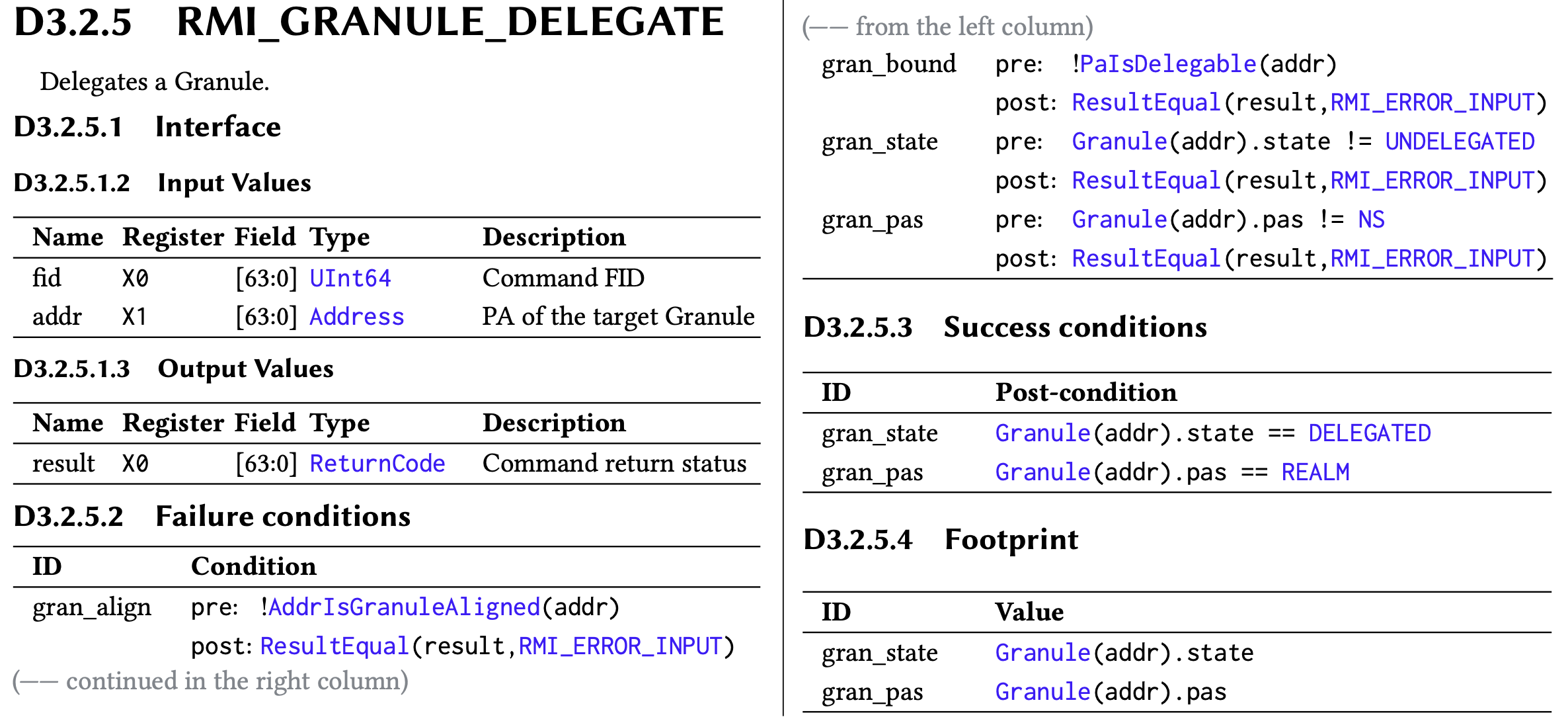}
  \caption{RMI\_GRANULE\_DELEGATE command specification (draft)~\cite{fox2023verification}}
  \label{fig:spe}
\end{figure}

Fig.~\ref{fig:spe} illustrates an example of RMM specification of the RMI\_GRANULE\_D ELEGATE command.\footnote{This is an example from the drafted specification version accessed on early 2022~\cite{tfrmm}.}
The specification takes a single granule address as input from the Host (the \textit{addr} parameter, in general purpose
register \textit{X1}, is the physical address location of the granule to be delegated). If the corresponding granule metadata is UNDELEGATED, meaning that the granule is not in the REALM PAS, and it is currently in the NS PAS, then the granule is moved to the REALM PAS. The failure conditions consist of several pairs of pre/post conditions, each of which is a possible error condition. When no failure occurs, the success conditions describe the expected updates to the RMM machine state.

%---------------------------------------------------------
\section{RMM Verification with ESBMC}
%---------------------------------------------------------

Listing~\ref{lst:impl} illustrates a C example of a verification harness generated from the RMM specification. It initializes a non-deterministic global state from line $1$ to $7$. In line $9$, it executes the actual command of the specification. Line $8$ and line $10$ evaluate the pre- and post-conditions using the Boolean variables \textit{failure\_src\_align\_pre} and \textit{failure\_src\_align\_post}, respectively. Lines 11-16 are the boolean conditions of the pre- and post-conditions. Finally, the assertion in line 17 checks the post-condition if the pre-condition fails. From this verification harness, we can use any off-the-shelf software model checker to automatically check the assertion for the post-condition. Properly choosing the underlying verification strategy and parameters to configure ESBMC is essential to achieving efficient verification of RMM. Here, we introduce two main verification techniques available in ESBMC: \textit{bounded verification} to unwind each loop occurring in the program with a different upper bound and \textit{multi-property check} to verify all properties incrementally using the underlying SMT solver.

 \begin{lstlisting}[language=C, basicstyle=\scriptsize\ttfamily, numbers=left, numberstyle=\scriptsize, frame=single, breaklines=true, captionpos=b, caption={Verification harness from the specification~\cite{fox2023verification}}, label=lst:impl]
struct tb_regs __tb_regs = __tb_arb_regs();
__tb_regs.X0 = SMC_RMM_DATA_CREATE;
__tb_regs.X1 = nondet_uint64_t(); // data
__tb_regs.X2 = nondet_uint64_t(); // rd
__tb_regs.X3 = nondet_uint64_t(); // map_addr
__tb_regs.X4 = nondet_uint64_t(); // src
__init_global_state(__tb_regs.X0); // Generate non-deterministic state
bool failure_src_align_pre = !AddrIsGranuleAligned(src); // Precondition 
uint64_t result = tb_handle_smc(&__tb_regs); // Execute command
bool failure_src_align_post = ResultEqual(result, RMI_ERROR_INPUT); // Postcondition
// Failure condition assertions (excerpt)
bool prop_failure_src_align_ante = failure_src_align_pre; 
__COVER(prop_failure_src_align_ante);
if (prop_failure_src_align_ante) {
bool prop_failure_src_align_cons = failure_src_align_post; 
__COVER(prop_failure_src_align_cons); 
__ASSERT(prop_failure_src_align_cons , "prop_failure_src_align_cons"); }
\end{lstlisting}

%----------------------------------------------------
\subsection{Bounded Verification}
\label{sec:BoundedConfiguration}
%----------------------------------------------------

We can configure the loops' unrolling depth for BMC tools to reduce the program state space for refutation/verification. In particular, we must limit the unwinding bound to avoid an unbounded loop being unrolled infinitely. However, we must carefully configure the bounds for complex programs such as with nested loops because the verification time could increase dramatically. Both ESBMC and CBMC support setting different upper bounds for each loop in the program. While this setup can be complicated compared to setting a unified bound for all loops or unrolling them incrementally, it can result in smaller state spaces and faster checking speeds, as shown in our experimental evaluation. This happens because iteratively unrolling loops and calling a solver via incremental and induction methods can be costly, especially for programs containing various nested loops. We refer the reader to our previous work about handling loops in BMC for a detailed discussion about this topic~\cite{AlhawiRGCF21, GadelhaIC17}, including comparisons to state-of-the-art verification methods. To verify the components of the Arm Confidential Computing Architecture, which is the focus of this paper, our preliminary experiments on incremental BMC have shown a significant performance difference when we set different upper bounds for each loop, which will be presented in Section~\ref{sec:experiments}. 

%-------------------------------------------------
\subsection{Multi-Property Check}
%-------------------------------------------------

Most real-world verification instances contain multiple assertions~\cite{abs-2311-05281}. This is because any non-trivial program must satisfy several invariants at once. However, multiple assertions create a crucial dilemma for software verification tools: whether we encode them all in a single large formula or split them into separate smaller ones. A single large formula is faster to solve than smaller ones because the latter have multiple (or incremental) queries to a solver~\cite{MenezesAFLMSSBGTKC24}. However, it also raises the problem that the counterexample in a single formula can only expose one property violation in the program.

By default, ESBMC runs in single property check mode, which encodes all verification conditions (VCs) into one single SMT formula and terminates once if it finds a violation by the underlying SMT solver across the program path. ESBMC uses Boolector as its default SMT solver~\cite{NiemetzPB14}. Recently, ESBMC added verification support for multi-property checks (see Listing~\ref{lst:mul}) to report all property violations for a single call. During the multi-property check, each property assumes the other properties are unreachable by keeping the current encoded property. In the example of Listing~\ref{lst:mul}, none of the two assertions hold because the value of $a$ is non-deterministic and can reach both cases to trigger the assertion. ESBMC, in single property check mode, will terminate once it finds a violation, e.g., \verb|assert(a>1)|, while multi-property check mode aims at reporting both of these violations.

    \begin{lstlisting}[language=C, basicstyle=\scriptsize\ttfamily, frame=single, breaklines=true, captionpos=b, caption={A program with two property violations.}, label=lst:mul]
#include <assert.h>
extern int nondet_int();
int main() {
  int a = nondet_int();
  switch (a) {
  case 0: assert(a > 0); break;
  case 1: assert(a > 1); break;
  default: return 0;
  }
}
    \end{lstlisting}

%----------------------------------------------  
\section{Experiment Results}
\label{sec:experiments}
%----------------------------------------------

In this section, we evaluate the capability of ESBMC single- and multi-property checks on RMM.

\paragraph{Experimental Setup.} We conduct the experiments on a Ubuntu OS with $16$GB memory and an Intel i$7$ processor, limiting the timeout to $5000$\,s. The benchmarks are listed in Table~\ref{tab:res} and are open-source~\cite{tfrmm}. We compare the results between ESBMC v7.4 and CBMC v5.94. Unless we state otherwise, we run ESBMC with the flag \texttt{--unwindset}. This flag ensures it uses the same default approach of CBMC, i.e., unrolling each loop to the maximum number of loop iterations occurring in the program without producing under-approximation. ESBMC adds explicit assertions to check whether the loop is fully unrolled.

\paragraph{Number of Violations Found.} Table~\ref{tab:res} shows the total number of failures found by ESBMC and CBMC. The main finding is that ESBMC reports more failures than CBMC in three test cases. 
\vspace{-2ex}
\paragraph{\textcolor{black}{Confirmed and Unconfirmed Violations.}} We are still working on confirming every bug reported by ESBMC due to the complexity and readability of the output traces. In particular, ESBMC provides traces that describe hundreds or thousands of states, where each state contains information about the program location and the value of the local/global variables. Still, we confirmed a correct positive in the test case \verb|RMI_REALM_DESTROY|, which is caused by converting a pointer to an integer and can affect the global verification conditions in that test case. The actual RMM code was fixed, though waiting for code review, before we conducted this experiment, as it involved undefined behavior due to that conversion. We present a characterized example in Listing~\ref{lst:ptoi}, and we have reported the issue to the CBMC team so that they could improve their verifier. A CBMC developer has confirmed the bug in their memory model (see \url{https://github.com/diffblue/cbmc/issues/8161}).
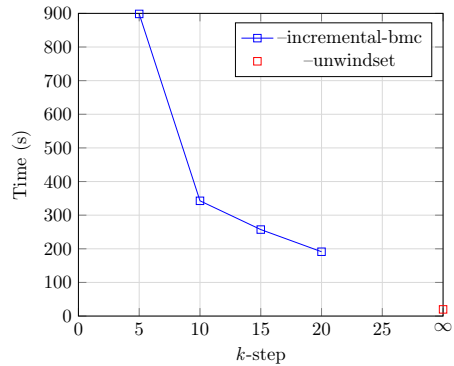
\begin{figure}[t]
\centering
\resizebox{0.5\textwidth}{!}{%
\begin{tikzpicture}
\begin{axis}[
    xlabel={\textit{k}-step},
    ylabel={Time (s)},
    xmin=0, xmax=30, 
    ymin=0, ymax=900, 
    xtick={0,5,10,15,20,25,30},
    xticklabels={0,5,10,15,20,25,$\infty$},
    ytick={0,100,200,300,400,500,600,700,800,900},
    legend pos=north east,
    grid=major,
    grid style={line width=.1pt, draw=gray!30}
]

\addplot[
    color=blue,
    mark=square,
    ]
    coordinates {
    (5,898.4)(10,342.7)(15,257.2)(20,191.3)
    };
    \legend{--incremental-bmc, --unwindset}

\addplot[
    color=red,
    mark=square,
    mark options={fill=none},
    only marks,
    ]
    coordinates {
    (30,19.8)
    };
    
\end{axis}
\end{tikzpicture}
}
\caption{Verification time for different \textit{k}-steps during each loop rolling. The \textbf{red} node at the bottom left represents unrolling the loops completely until the given bounds.
}
\label{fig:incr}
\end{figure}
\vspace{-2ex}
\paragraph{Impact of step $k$ in incremental BMC.} ESBMC provides an option to set the $k$-step when incrementally unrolling the loops. For example, a $k$-step of $5$ will unfold a given loop incrementally as $5$, $10$, $15$, and $20$ (maximum $k$-step). We conduct an extra experiment on a single but representative RMM test case to check the impact of step $k$ in incremental BMC. In Fig.~\ref{fig:incr}, it can be seen that although increasing the \textit{k}-steps can reduce looping unrolling and solving time since we can find a property violation faster, it is not as efficient as ESBMC non-iterative mode, i.e., via unwindset that unrolls the loops directly to the upper bound. These results explain our chosen methodology as described in Section~\ref{sec:BoundedConfiguration}. 

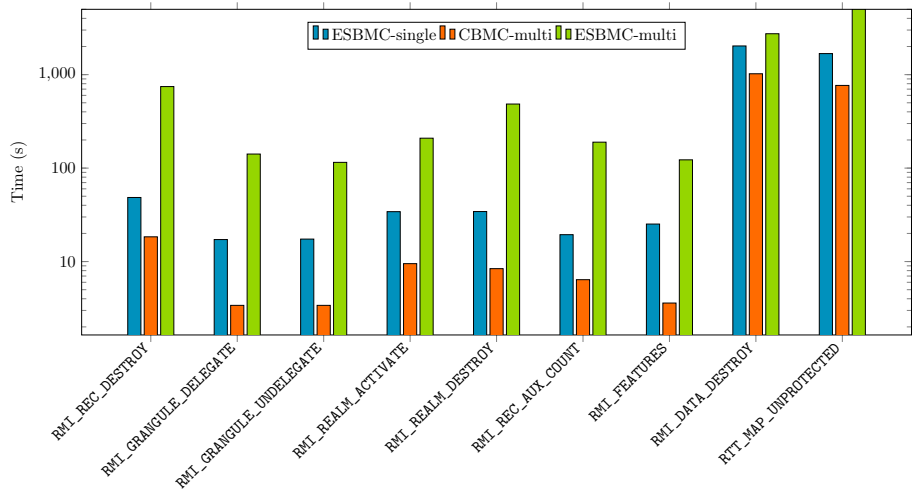
\begin{figure}[t]
  \centering
  \resizebox{\textwidth}{!}{
    \begin{tikzpicture}
      \begin{axis}[
        ybar,
        width=1.0\textheight,
        height=0.7\textwidth,
        symbolic x coords={RMI\_REC\_DESTROY,RMI\_GRANGULE\_DELEGATE,RMI\_GRANGULE\_UNDELEGATE,RMI\_REALM\_ACTIVATE,RMI\_REALM\_DESTROY,RMI\_REC\_AUX\_COUNT,RMI\_FEATURES,RMI\_DATA\_DESTROY,RTT\_MAP\_UNPROTECTED},
        xticklabels={\texttt{RMI\_REC\_DESTROY},\texttt{RMI\_GRANGULE\_DELEGATE},\texttt{RMI\_GRANGULE\_UNDELEGATE},\texttt{RMI\_REALM\_ACTIVATE},\texttt{RMI\_REALM\_DESTROY},\texttt{RMI\_REC\_AUX\_COUNT},\texttt{RMI\_FEATURES},\texttt{RMI\_DATA\_DESTROY},\texttt{RTT\_MAP\_UNPROTECTED}},
        xtick=data,
        x tick label style={rotate=45, anchor=east}, 
        ylabel={Time (s)},
        bar width=8pt,
        legend style={at={(0.5,-0.15)},anchor=north,legend columns=-1, yshift=220pt},
        ymode=log,
        log ticks with fixed point, 
        ymax = 5000,
        ]
        
        % Data set 1
        \addplot[fill=arm-blue] coordinates {(RMI\_REC\_DESTROY,48.5) (RMI\_GRANGULE\_DELEGATE,17.2) (RMI\_GRANGULE\_UNDELEGATE,17.4) (RMI\_REALM\_ACTIVATE,34.2) (RMI\_REALM\_DESTROY,34.3) (RMI\_REC\_AUX\_COUNT,19.4) (RMI\_FEATURES,25.2) (RMI\_DATA\_DESTROY,2027.4) (RTT\_MAP\_UNPROTECTED,1680.1)};
        
        % Data set 2
        \addplot[fill=arm-orange] coordinates {(RMI\_REC\_DESTROY,18.4) (RMI\_GRANGULE\_DELEGATE,3.4) (RMI\_GRANGULE\_UNDELEGATE,3.4) (RMI\_REALM\_ACTIVATE,9.5) (RMI\_REALM\_DESTROY,8.4) (RMI\_REC\_AUX\_COUNT,6.4) (RMI\_FEATURES,3.6) (RMI\_DATA\_DESTROY,1022) (RTT\_MAP\_UNPROTECTED,766.9)};
        
        % Data set 3
        \addplot[fill=arm-green] coordinates {(RMI\_REC\_DESTROY,745.5) (RMI\_GRANGULE\_DELEGATE,141.4) (RMI\_GRANGULE\_UNDELEGATE,115.2) (RMI\_REALM\_ACTIVATE,209.0) (RMI\_REALM\_DESTROY,484.7) (RMI\_REC\_AUX\_COUNT,189.5) (RMI\_FEATURES,122.6) (RMI\_DATA\_DESTROY,2739.7) (RTT\_MAP\_UNPROTECTED,5000)};
        
        \legend{ESBMC-single, CBMC-multi, ESBMC-multi}
      \end{axis}
    \end{tikzpicture}
  }
  \caption{Experiment result, where x-axis represents the RMM testcases, and y-axis represents the verification time in seconds. Different colors: blue on the left is for ESBMC single-property check, red on the middle is for CBMC multi-property check, and yellow on the right is for ESBMC multi-property check.}
  \label{fig:result}
\end{figure}

\begin{lstlisting}[language=C, basicstyle=\scriptsize\ttfamily, frame=single, breaklines=true, label=lst:ptoi, caption={Pointer to integer convertion example extracted from rmm, where the assertion should fail if the address falls into a negative value, for example, -2048.}, captionpos=b]
#include<assert.h>
int arr[8] = {1,2,3,4,5,6,7,8};
int main() {
  int *a = &arr[7];
  if((unsigned long)a >= (unsigned long)(-4095))
    assert((unsigned long)(-1*(long)a) < 6); //esbmc fails, cbmc  success
}
\end{lstlisting}
\begin{table}[t]
  \centering
  \caption{Verification results of multi-property checks using ESBMC and CBMC.
  }
  \label{tab:perfor}
  \label{tab:res}
  \small 
  \begin{tabular}{l@{\qquad}c c@{\quad\qquad}c c}
    \toprule
    \multirow{2}{*}{Command} & \multicolumn{2}{c@{\quad\qquad}}{Assert Fail} & \multicolumn{2}{c}{VCCs/Solver Calls} \\
    & ESBMC & CBMC & ESBMC & CBMC \\
    \midrule
    \verb|RMI_REC_DESTROY| & 20 & 20 & 113/113 & 142/19 \\
    \verb|RMI_GRANULE_DELEGATE| & safe & safe & 54/54 & 132/2\\
    \verb|RMI_GRANULE_UNDELEGATE| & 1 & 1 & 45/45 & 132/1\\
    \verb|RMI_REALM_ACTIVATE| & \textbf{3} & \textbf{safe} & 53/53 & 140/1\\
    \verb|RMI_REALM_DESTROY| & \textbf{17} & \textbf{1} & 114/114 & 148/2\\
    \verb|RMI_REC_AUX_COUNT| & 1 & 1 & 48/48 & 139/2\\
    \verb|RMI_FEATURES| & safe & safe & 21/21 & 125/1\\
    \verb|RMI_DATA_DESTROY| & \textbf{>=26} & \textbf{22} & 82/82 & 151/18\\
    \bottomrule
  \end{tabular}
\end{table}
\vspace{-3ex}
\paragraph{Performance Comparison.} Fig.~\ref{fig:result} illustrates the verification time of the test cases for three runs: ESBMC single property check, CBMC multi-property check, and ESBMC multi-property check. Note that the default option of CBMC always checks all properties in one execution, which is relevant when verifying industrial code. CBMC and ESBMC single property checks can finish most verification tasks within a few seconds. However, the ESBMC multi-property check needs a few minutes for each task and even longer for the last case in the figure, which has exceeded the timeout of $5000$\,s. Table~\ref{tab:perfor} provides an insight into the performance difference. ESBMC naively calls an SMT solver when there is a generated SMT formula (per property to check), and it takes time to solve them individually. Meanwhile, CBMC can exploit the power of incremental SAT solving by MiniSAT to reuse solved instances for new problems~\cite{KroeningS16}. Another important factor is that CBMC implements a string solver to solve frequent string operations at RMM code, while ESBMC has not implemented one yet, which results in a performance slowdown for strings.

\vspace{-2ex}
\paragraph{Syntax Errors.} In addition to the correctness and performance, there exists an interesting aspect related to the deployed front-end parsers: CBMC uses a modified C parser, while ESBMC implements Clang API to transform the source code into Clang AST~\cite{GadelhaMMC0N18}, without having details of the input program compiled away. The latter can do strict and elegant code syntax checks before starting model checking. ESBMC can also provide compilation error messages as expected from a compiler and leverage Clang’s static analyzer to provide meaningful warnings when parsing the program. For example, consider Listing ~\ref{lst:warn} that illustrates an example from the RMM implementation, where a declaration of \textbf{struct} object is in the switch case but without curly braces. ESBMC reports the error using Clang, while the C parser in CBMC ignores it. 

\begin{lstlisting}[language=C, basicstyle=\scriptsize\ttfamily, frame=single, breaklines=true, captionpos=b, caption={Code segement did not compile for Clang since it cannot declare variables inside a case block without enclosing it in braces.}, label=lst:warn]
...
case SMC_RMM_RTT_READ_ENTRY:
    struct smc_result rst; 
    smc_rtt_read_entry(*X1, *X2, *X3, &rst);
    result = rst.x[0]; *X1 = rst.x[1]; *X2 = rst.x[2]; 
    *X3 = rst.x[3]; *X4 = rst.x[4];
    break;
...
\end{lstlisting}

%----------------------------------------  
\section{Conclusions}
%----------------------------------------

We present our application of ESBMC to verify newly introduced RMM components in Arm CCA. On the one hand, we show that to ensure firmware correctness, we can fully verify it by applying various techniques, each with its strengths and weaknesses. In our experiment, we applied ESBMC and found more violation properties than CBMC. On the other hand, we show the community that we have to write more scalable verifiers to check large and complex code bases for multi-property checks. In addition, we recommend that software verifiers connect their verification algorithms with industrial-strength compilers, as we find that self-developed parsers may not be precise enough. Lastly, we should not only focus on the verification algorithms but also try to produce more readable verification results to promote the power of formal verification techniques in the software industry. Future work for ESBMC includes more efficient algorithms for faster SMT solving for multiple property verification and checking data races in multi-threaded programs. %In our evaluation, we could not check for data races because current state-of-the-art software verifiers are not yet ready to cope with the RMM concurrency safety properties within a reasonable budget since they involve non-deterministic data and thread interleavings. 

%---------------------------------------------
\section*{Acknowledgments}
%---------------------------------------------

The work in this paper is partially funded by the Arm Center of Excellence at the University of Manchester, UK, EPSRC grants EP/T026995/1, EP/V000497/1, EU H2020 ELEGANT 957286, and Soteria project awarded by the UK Research and Innovation for the Digital Security by Design (DSbD) Programme. We acknowledge Franz Brauße and Chenfeng Wei, who helped refine ESBMC for RMM verification. We would also like to thank engineers from \texttt{tf-rmm} team (\url{https://www.trustedfirmware.org/projects/tf-rmm/}). 

\appendix
\section{Appendix}
This section lists the details from above results where ESBMC reports more failures from RMM testcases than CBMC.

\begin{table}[ht]
\centering
\caption{Unique results from ESBMC}
\scriptsize
\begin{tabular}{ l l l }
\toprule
No. & Location & Description \\
\midrule
1 & \texttt{tb\_rmi\_realm\_activate.c} line 107 & \texttt{prop\_success\_realm\_state\_cons} \\
1 & \texttt{tb\_rmi\_realm\_activate.c} line 98 & \texttt{prop\_result\_cons} \\
1 & \texttt{tb\_rmi\_realm\_activate.c} line 89 & \texttt{prop\_failure\_realm\_state\_cons} \\
2 & \texttt{tb\_rmi\_realm\_destroy.c} line 123 & \texttt{prop\_success\_rd\_state\_cons} \\
2 & \texttt{tb\_rmi\_realm\_destroy.c} line 115 & \texttt{prop\_success\_rtt\_state\_cons} \\
2 & \texttt{tb\_rmi\_realm\_destroy.c} line 106 & \texttt{prop\_result\_cons} \\
2 & \texttt{granule.h} line 70 & assertion \texttt{false} \\
2 & \texttt{granule.h} line 66 & assertion \texttt{g->refcount == 0UL} \\
2 & \texttt{granule.h} line 63 & assertion \texttt{g->refcount <= GRANULE\_SIZE / sizeof(uint64\_t)} \\
2 & \texttt{granule.h} line 59 & assertion \texttt{g->refcount == 0UL} \\
2 & \texttt{granule.h} line 56 & assertion \texttt{granule\_refcount\_read\_relaxed(g) <= 1UL} \\
2 & \texttt{granule.h} line 46 & assertion \texttt{g->refcount == 0UL} \\
2 & \texttt{granule.h} line 43 & assertion \texttt{granule\_refcount\_read\_relaxed(g) == 0UL} \\
2 & \texttt{tb\_lock.c} line 80 & The granule must be locked. \\
2 & \texttt{granule.h} line 120 & assertion locked \\
2 & \texttt{tb\_lock.c} line 49 & The granule lock must be free. \\
2 & \texttt{status.h} line 90 & assertion \texttt{(unsigned int)(-1 * (long)ptr) < RMI\_ERROR\_COUNT} \\
2 & \texttt{vmid.c} line 63 & assertion \texttt{vmid < vmid\_count} \\
2 & \texttt{realm.c} line 337 & unwinding assertion loop \\
3 & \texttt{granule.h} line 120 & assertion locked \\
3 & \texttt{granule.c} line 47 & assertion \texttt{idx < RMM\_MAX\_GRANULES} \\
3 & \texttt{tb\_lock.c} line 49 & The granule lock must be free. \\
3 & \texttt{s2tt.c} line 358 & assertion \texttt{map\_addr < (1UL << ipa\_bits)} \\
3 & \texttt{s2tt.c} line 357 & assertion \texttt{level >= start\_level} \\
3 & \texttt{s2tt.c} line 356 & assertion \texttt{start\_level >= MIN\_STARTING\_LEVEL} \\
\bottomrule
\end{tabular}
\label{tab:assertions}
\end{table}
\printbibliography
\end{document}

%% file: cca-diagram.tex
\begin{tikzpicture}[font=\scriptsize]
% all the coordination is relative scriptsize the left bottom cornor, i.e. EL-3 node.
\node (el3) {\texttt{EL-3}};
\node (el2) at ($(el3.north) + (0,0.6)$) {\texttt{EL-2}};
\node (el1) at ($(el2.north) + (0,0.7)$) {\texttt{EL-1}};
\node (el0) at ($(el1.north) + (0,0.3)$) {\texttt{EL-0}};

%el-3
\node[anchor=west] (root) at ($(el3.east) + (0.8,0)$) {\textbf{Root}};
\node[fill=arm-blue, anchor=west, minimum width=260pt] (monitor) at ($(el3.east) + (2.3,0)$) {Monitor};

%el-2
\node[fill=arm-orange, anchor=west, minimum width=99pt, align=center] (rmm) at ($(el2.east) + (0.2,0)$) {
    \begin{tabular}{@{}c@{}}
        Realm Management \\
        Monitor (RMM)
    \end{tabular}
};
\node[fill=arm-dark-blue, text=white, anchor=west, minimum width=92pt, align=center] (hypervisor) at ($(rmm.east) + (0.3,0)$) {Hypervisor};
\node[fill=arm-green!80!black, anchor=west, minimum width=110pt, align=center] (spm) at ($(hypervisor.east) + (0.33,0)$) {
    \begin{tabular}{@{}c@{}}
        Secure Partition \\
        Manager (SPM)
    \end{tabular}
};
%el-1 and 0
\node[fill=arm-yellow, anchor=west, minimum width=40pt] (kernel1) at ($(el1.east) + (0.35,0)$) {Kernel};
\node[fill=arm-yellow, anchor=west, minimum width=40pt] (kernel2) at ($(kernel1.east) + (0.35,0)$) {Kernel};
\node[fill=arm-yellow, anchor=west, minimum width=40pt] (realmapp1) at ($(el0.east) + (0.35,0)$) {App};
\node[fill=arm-yellow, anchor=west, minimum width=40pt] (realmapp2) at ($(realmapp1.east) + (0.35,0)$) {App};
\node[draw, arm-orange, line width=1pt, fit=(kernel1)(realmapp1)] (realm1) {};
\node[draw, arm-orange, line width=1pt, fit=(kernel2)(realmapp2)] (realm2) {};
\node[draw, arm-orange, dashed, line width=1pt, fill=arm-light-gray, minimum width=30pt, rotate=90] (rvm1) at ($(kernel2.north east)!0.5!(realmapp2.south east) + (0.6,0)$) {RVM};
\node[draw, arm-orange, dashed, line width=1pt, fill=arm-light-gray, minimum width=30pt, rotate=90] (rvm2) at ($(rvm1.south) + (0.4,0)$) {RVM};

\node[fill=arm-dark-gray, minimum width=30pt, rotate=90] (vm) at ($(rvm2.south) + (0.4,0)$) {VM};
\node[fill=arm-dark-gray, anchor=west, minimum width=40pt, align=center] (os) at ($(kernel2.east) + (2.2,-0.2)$) {OS Kernel};
\node[fill=arm-green!90!black, anchor=west, minimum width=40pt,align=center] (tos) at ($(os.east) + (0.2,0)$) {Trusted OS};
\node[fill=arm-green!90!black, anchor=west, minimum width=40pt,align=center] (securepar) at ($(tos.east) + (0.08,0)$) {Secure Partition};
\node[fill=arm-light-gray, rotate=90, minimum width=18pt] (app1) at ($(realmapp2.east) + (2.5,0)$) {App};
\node[fill=arm-light-gray, rotate=90, minimum width=18pt] (app2) at ($(app1.south) + (0.5,0)$) {App};
\node[fill=arm-green, rotate=90, minimum width=18pt] (ta1) at ($(app2.south) + (0.8,0)$) {TA};
\node[fill=arm-green, rotate=90, minimum width=18pt] (ta2) at ($(ta1.south) + (0.7,0)$) {TA};
\node[anchor=west, inner sep=0pt] (realm) at ($(realm1.north west) + (0,0.5)$) {\textbf{Realm}};
\node[anchor=west, inner sep=0pt] (nonsercure) at ($(rvm1.north east) + (0,0.5)$) {\textbf{Non-secure}};
\node[anchor=west, inner sep=0pt] (sercure) at ($(ta1.north east) + (0,0.5)$) {\textbf{Secure}};

% arrow
\draw[-latex, line width=1pt] (rvm1.east) to[bend right=20] (realm1.north);
\draw[-latex, line width=1pt] (rvm2.east) to[bend right=30] (realm2.north);

% separation
\coordinate (el-2-3-y) at ($(rmm.south)!0.5!(root.north)$);
\draw[-, dashed, line width=1pt] (el-2-3-y -| rmm.west) -- (el-2-3-y -| spm.east);

\coordinate (realm-normal-x) at ($(rmm.east)!0.5!(hypervisor.west)$);
\coordinate (el-0-top-y) at ($(realm2.north east)!0.5!(rvm1.north east)$);
\draw[-, dashed, line width=1pt] (el-2-3-y -| realm-normal-x) -- (el-0-top-y -| realm-normal-x);

\coordinate (normal-tz-x) at ($(hypervisor.east)!0.5!(spm.west)$);
\coordinate (normal-tz-top-y) at ($(app2.south east)!0.5!(ta1.north east)$);
\draw[-, dashed, line width=1pt] (el-2-3-y -| normal-tz-x) -- (el-0-top-y -| normal-tz-x);

\coordinate (el-1-2-y) at ($(rmm.north)!0.5!(realm1.south)$);
\draw[-, dashed, line width=0.5pt] (el-1-2-y -| rmm.west) -- (el-1-2-y -| rmm.east);

\coordinate (realm-realm-x) at ($(realm1.east)!0.5!(realm2.west)$);
\draw[-, dashed, line width=0.5pt] (el-1-2-y -| realm-realm-x) -- (el-0-top-y -| realm-realm-x);

\draw[-, dashed, line width=0.5pt] (el-1-2-y -| rvm1.north) -- (el-1-2-y -| os.east);

\coordinate (rvm-rvm-x) at ($(rvm1.south)!0.5!(rvm2.north)$);
\draw[-, dashed, line width=0.5pt] (el-1-2-y -| rvm-rvm-x) -- (el-0-top-y -| rvm-rvm-x);

\coordinate (rvm-vm-x) at ($(rvm2.south)!0.5!(vm.north)$);
\draw[-, dashed, line width=0.5pt] (el-1-2-y -| rvm-vm-x) -- (el-0-top-y -| rvm-vm-x);

\coordinate (vm-os-x) at ($(vm.south)!0.5!(os.west)$);
\draw[-, dashed, line width=0.5pt] (el-1-2-y -| vm-os-x) -- (el-0-top-y -| vm-os-x);

\coordinate (el-0-1-y) at ($(os.north)!0.5!(app1.west)$);
\draw[-, dashed, line width=0.5pt] (el-0-1-y -| os.north west) -- (el-0-1-y -| os.north east);

\coordinate (app-app-x) at ($(app1.south)!0.5!(app2.north)$);
\draw[-, dashed, line width=0.5pt] (el-0-1-y -| app-app-x) -- (el-0-top-y -| app-app-x);

\draw[-, dashed, line width=0.5pt] (el-1-2-y -| tos.west) -- (el-1-2-y -| securepar.east);

\coordinate (tos-par-x) at ($(tos.east)!0.5!(securepar.west)$);
\draw[-, dashed, line width=0.5pt] (el-1-2-y -| tos-par-x) -- (el-0-top-y -| tos-par-x);

\draw[-, dashed, line width=0.5pt] (el-0-1-y -| tos.north west) -- (el-0-1-y -| tos.north east);

\coordinate (ta-ta-x) at ($(ta1.south)!0.5!(ta2.north)$);
\draw[-, dashed, line width=0.5pt] (el-0-1-y -| ta-ta-x) -- (el-0-top-y -| ta-ta-x);

\end{tikzpicture}